\documentclass[twocolumn,showpacs,preprintnumbers,amsmath,amssymb]{revtex4}

\usepackage{graphicx}
\usepackage{dcolumn}
\usepackage{bm}
\begin{document}
\title{Experimental realization of a low-noise heralded single photon source}
\author{G. Brida$^1$, I.P. Degiovanni$^1$, M. Genovese$^1$, A. Migdall$^2$, F.
Piacentini$^1$, S.V. Polyakov$^2$, I. Ruo Berchera$^1$} \affiliation{$^1$ I.N.RI.M., Strada delle
Cacce 91, 10135 Torino, Italia} \affiliation{$^2$ Joint Quantum Institute and National Institute of
Standards and Technology, 100 Bureau Dr, Stop 8441, Gaithersburg, MD, 20899 USA}
\begin{abstract}
We present a heralded single-photon source with a much lower level of unwanted background photons
in the output channel by using the herald photon to control a shutter in the heralded channel. The
shutter is implemented  using a simple field programable gate array controlled optical switch.
\end{abstract}
\maketitle

While ideal single-photon sources are desired for many applications from metrology
\cite{metrology}, to quantum information \cite{quantumcomm,quantumtech}, to analytical methods, to
foundations of quantum mechanics \cite{alicki} the best that can be achieved are sources that offer
some approximation to such a source.

One commonly used approximation is the heralded-photon source which relies on photons produced in
pairs, where one of the photons is used to herald the existence of the other photon. While a useful
device, this type of source suffers from two particular deficiencies, and these deficiencies have
afflicted pair sources from the earliest pair sources based on atomic cascade \cite{ACT}, to
parametric down conversion (PDC) \cite{PDC1,PDC2,PDCwg1,PDCwg2} in crystals, to four-wave mixing in
fibers \cite{4WM2,4WM3,4WM4}. One deficiency is that the production is probabilistic and the other
is that the probability of extraction of each of the photons of a pair is independent and less than
unity. This second deficiency results in many heralding counts that yield no output photon and
conversely many photons are emitted from the output channel without a heralding count. Both of
these failure modes can present problems for particular applications and are worthy of efforts to
reduce their likelihood. Both deficiencies can be reduced by improving the photon extraction
efficiency and there are efforts in that direction \cite{coupling1,PDC2,coupling2}. To obtain a
further reduction in the emission of unheralded photons beyond improving the extraction efficiency,
several strategies have been proposed and implemented. For example, using a photon-number-resolving
detector on the heralding arm highlights the presence of multi-photon emission from the heralded
arm \cite{PNRD}. Another approach exploits the use of an optical shutter, where the optical output
path is blocked unless a photon is known to be incident. This simple idea has been discussed for
some time \cite{pt}, but source development efforts have been focused more on the production of
single-photon sources ``on-demand" \cite{ondemand}, rather than on the suppression of unheralded
photons. This shuttered or \textit{low-noise} heralded single photon source is particularly
advantageous when dealing with detectors with high temporal jitter or slow temporal response (e.g.,
transition edge superconducting microbolometers), where the low time resolution does not allow for
tight time discrimination between the desired heralded photons and unwanted background photons. To
compensate for this, data rates must often be reduced to very low levels.

We note that, reducing the noise of a heralded photon source would be of particular advantage in applications such as radiometry
where knowing the number of emitted photons is key to the
measurements. Noisy photon sources are also problematic for
quantum information applications where additional unwanted photons
make the already difficult task of processing a fragile quantum
state that much more difficult.

In this work we present a heralded-photon source
based on PDC with an optical shutter that opens for a short period
of time around the expected emission time of a heralded photon.
Because this scheme greatly reduces the emission of unheralded
photons, we refer to this type of source as a low-noise heralded-single-photon-source (HSPS). Despite the fact that
significant further improvements in
single-photon performance could be obtained, the current version of our source
is already at the level of the best solid-state based
single-photon source \cite{shell}.

\begin{figure}[tbh]
\includegraphics[width=55mm,angle=270]{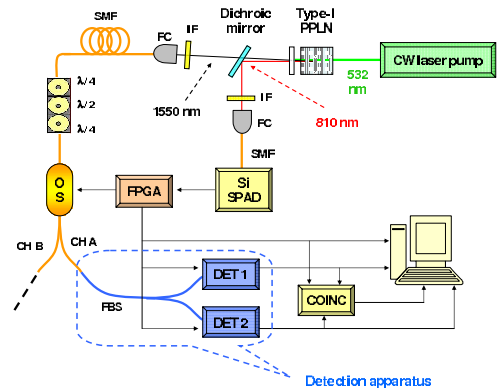}
\caption{Experiment arrangement. Channel A is the low-noise HSPS output. Channel B is sent to a
beam dump. } \label{f:setups}
\end{figure}

 In our experimental setup
(Fig.\ref{f:setups}) a continuous wave (cw) laser ($\lambda=532$ nm)
pumps a $5\times1\times5$ mm periodically poled Lithium Niobate (PPLN) crystal, producing non-degenerate
parametric down conversion
signal and idler photons with wavelengths of $\lambda_{s}=1550$ nm and $\lambda_{i}=810$ nm.\\
The idler photon is sent to an interference filter (IF) with a full width half maximum (FWHM) of 10
nm, then fiber-coupled to a silicon single-photon avalanche detector (Si-SPAD). The signal photon
is addressed to a 30 nm FWHM filter (IF) and coupled to a 20 m long single-mode optical fiber
connected to the optical switch (OS) controlled by a field programmable gate array (FPGA). The OS
channel A, chosen as our low-noise HSPS output channel, is connected to a 50\%-50\% fiber beam
splitter (FBS) whose outputs are sent to two infrared InGaAs SPADs ($\textrm{DET}_1$ and
$\textrm{DET}_2$), triggered by the same FPGA signal that triggers the optical switch. The InGaAs
SPAD detection window is 100 ns long. The outputs of the two InGaAs SPADs are sent to
the coincidence electronics and finally recorded by the computer.\\
The FPGA opens OS channel A for a time
interval $\Delta t_{\textrm{switch}}$ of only a few nanoseconds in coincidence with the detection of
an 810 nm photon, and then switches to channel B for a chosen
``shuttered'' time $t_{\textrm{dead}}$ before the system is ready to be
retriggered by a Si-SPAD count.\\
To reject InGaAs SPAD afterpulses, we set
$t_{\textrm{dead}}=20$ $\mu$s. We note that the minimum time step achievable by
our FPGA is $t_{\textrm{dead}}=6$ ns, thus, we are far from the performance limits of this technology.\\
We made measurements with four different switch pulse durations
$\Delta t_{\textrm{switch}}$ (60 ns, 30 ns, 15 ns, and 5 ns).
\begin{figure}[tbh]
\includegraphics[width=80mm,angle=0]{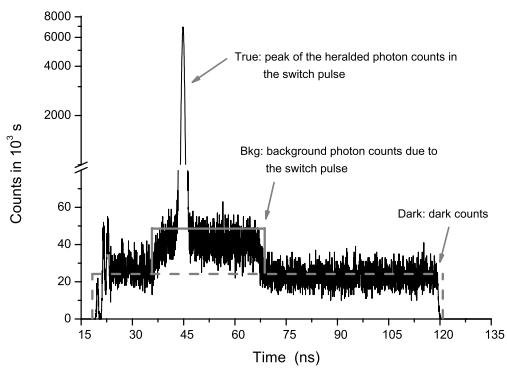}
\caption{Histogram of $\textrm{DET}_1$ detection window, with the peak inside the switch pulse
region ($\Delta t_{\textrm{switch}}=30$ ns); the true, background, and dark count contributions can
be clearly seen. } \label{peak-in}
\end{figure}

\begin{figure}[tbh]
\includegraphics[width=80mm,angle=0]{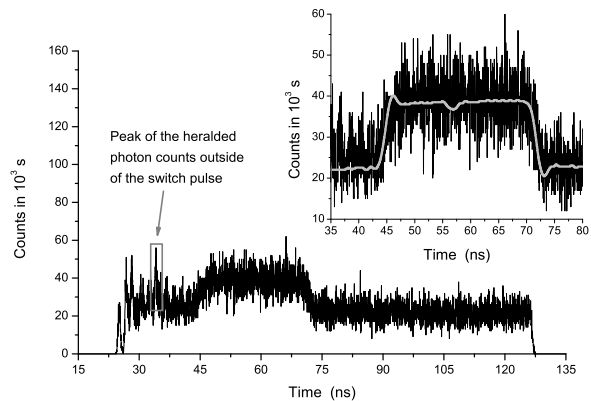}
\caption{Histogram of $\textrm{DET}_1$ detection window ($\Delta t_{\textrm{switch}}=30$ ns) when
the heralded photons peak is outside of the OS active region, therefore being highly suppressed.
The inset shows the switch-on region, with the solid line showing the shape of the electrical pulse
driving the OS. } \label{peak-out}
\end{figure}
Looking at the multichannel picture of the detection window of
$\textrm{DET}_1$ (Fig.\ref{peak-in}) we can distinguish three different
``regions'' corresponding to:
\begin{itemize}
    \item $N^\mathrm{(True)}$ = true heralded photon counts;
    \item $N^\mathrm{(Bkg)}$ = counts due to background and stray light passing
    through the optical switch;
    \item $N^\mathrm{(Dark)}$ = dark counts of the IR detector.
\end{itemize}
We define the true heralded photon detection probability for each trigger count as
    \begin{equation}\label{P-T}
    P_{i}^{\mathrm{(True)}}= \frac{N^{\mathrm{(True)}}}{N_{i}^{\mathrm{(Trig)}}}~~~~~~~~~~~~~~i=1,2
    \end{equation}
($P_{i}^{\mathrm{(Bkg)}}$ and $P_{i}^{\mathrm{(Dark)}}$ are analogously defined), where
$N_{i}^{\mathrm{(Trig)}}$ is the total number of trigger counts accepted by the $i$-th detector.\\
The overall detection probability of detector $i$ is
    \begin{equation}\label{P-sum}
    P_{i}^{\mathrm{(Tot)}}=P_{i}^{\mathrm{(True)}}+P_{i}^{\mathrm{(Bkg)}}+P_{i}^{\mathrm{(Dark)}}. 
    \end{equation}
To evaluate these three probabilities, we look at the histogramed outputs of $\textrm{DET}_1$ and
$\textrm{DET}_2$ in two different configurations: \textit{peak-in} (Fig.\ref{peak-in}), with the
heralded photons arriving in correspondence of the OS active region, and \textit{peak-out}
(Fig.\ref{peak-out}), where the switching pulse is delayed with respect to the arrival of the
heralded photons so that they do not arrive during the switch open time (i.e. the pulse duration $\Delta t _{\mathrm{switch}}$).\\
We can then calculate the ratio of unwanted to total photons in our distribution channel: we call
this parameter \textit{Output Noise Factor} ($ONF$), defined as:
    \begin{equation}\label{CNOF}
    ONF=\frac{P_{1}^{\mathrm{(Bkg)}}+P_{2}^\mathrm{(Bkg)}}{P_{1}^{\mathrm{(True)}}+P_{1}^{\mathrm{(Bkg)}}+P_{2}^{\mathrm{(True)}}+P_{2}^{\mathrm{(Bkg)}}
    }.
    \end{equation}
The other figure of merit that we consider for our HSPS is $\alpha$ (analogous to the second order
correlation function $g^{(2)}(0)$ \cite{grangier}):
    \begin{equation}\label{alfa}
    \alpha=\frac{P_{12}^{\mathrm{(True+Bkg;True+Bkg)}}}{P_{1}^{\mathrm{(True+Bkg)}} \cdot P_{2}^{\mathrm{(True+Bkg)}}},
    \end{equation}
where $P_{12}^{\mathrm{(True+Bkg;True+Bkg)}}$ is the probability of a coincidence photon count
between $\textrm{DET}_1$ and $\textrm{DET}_2$ (dark counts subtracted). Assuming
$P_{12}^{\mathrm{(True;True)}}=0$ (there is only one heralded photon in the fiber beam splitter per heralding
count), and $P_{i}^{\mathrm{(Bkg)}}$ and $P_{i}^{\mathrm{(Dark)}}$ are independent, we obtain:
    \begin{equation}\label{P12}
    P_{12}^{\mathrm{(True+Bkg;True+Bkg)}}=P_{12}^{\mathrm{(Tot;Tot)}}-P_{12}^{\mathrm{(Dark;tot)}}-$$$$-
    P_{12}^{\mathrm{(Tot;Dark)}}+P_{12}^{\mathrm{(Dark;Dark)}}.
    \end{equation}

\begin{figure}[tbh]
\includegraphics[width=80mm,angle=0]{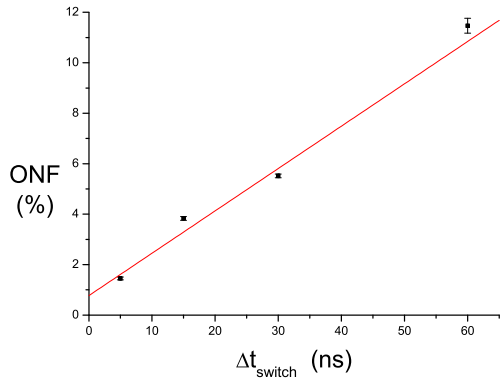}
\caption{$ONF$ parameter as a function of the switching time $\Delta t_{\textrm{switch}}$. The
linear fit (line) of the data (points) shows a correlation factor $R=0.985$. }  \label{ONF-plot}
\end{figure}

\begin{figure}[tbh]
\includegraphics[width=80mm,angle=0]{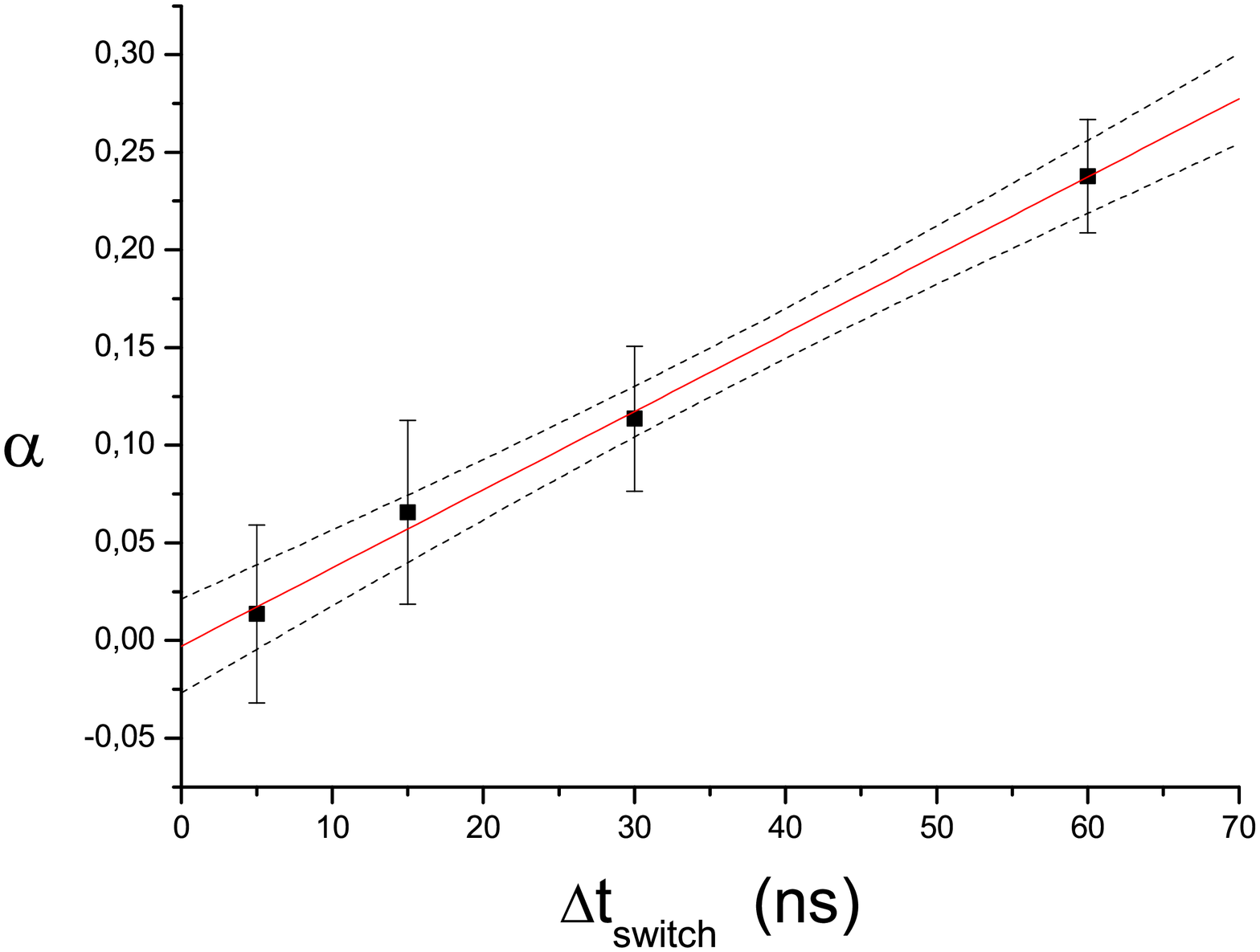}
\caption{Behavior of the parameter $\alpha$ ($\thickapprox g^{(2)}(0)$) as a function of the
switching time $\Delta t_{\textrm{switch}}$ (expressed in nanoseconds). The linear fit (solid line)
of the data (points) shows a correlation factor $R=0.998$, along with 95\% confidence bands (dashed
curves). \label{alfa-plot}}
\end{figure}

All the terms in Eq.(\ref{P12}) can be extracted from measurements
made by blocking the light to each detector in turn.\\
Our results, summarized in Figs. \ref{ONF-plot} and
\ref{alfa-plot}, show the $ONF$
decreasing linearly with the duration of $\Delta t_{\textrm{switch}}$, a
direct consequence of the reduction in background photons as the OS on-time is narrowed. The
values range from a maximum of $11.5\%$ for $\Delta t_{\textrm{switch}}=60$ ns
to a minimum of $1.45\%$ for $\Delta t_{\textrm{switch}}=5$ ns, clearly
showing the noise reduction in our source's output channel.

As expected, the parameter $\alpha$ shows the same behavior as the $ONF$, decreasing linearly with
the switching time $\Delta t_{\textrm{switch}}$: it ranges from 0.253 ($\Delta
t_{\textrm{switch}}=60$ ns) to the remarkable value 0.0136 ($\Delta t_{\textrm{switch}}=5$ ns),
highlighting the advantage of our shuttered single-photon source. In fact, our best measured
$\alpha$ value ($\alpha = 0.0136$) is comparable with, or even better than, the best values
obtained for single-photon emitters such as for example, a quantum dot in micropillar presenting
$\alpha = 0.02$ \cite{sps1} or $^{40}$Ca$^{+}$ in ion-trap cavity presenting $\alpha = 0.015 $
\cite{sps2}. Fig. \ref{alfa-plot} shows a linear fit to the data, where for the ideal case of
$\Delta t_{\mathrm{switch}} = 0$, we would obtain $\alpha = -0.003 \pm 0.025$, which is clearly
compatible with 0, indicating that there are no other effects limiting the device performance to
this level of uncertainty.
\\
Here, the uncertainties on the $\alpha$ data are larger than those obtained for the $ONF$, mainly
because the double coincidence events needed to evaluate $P_{12}^{\mathrm{(True+Bkg;True+Bkg)}}$
are relatively rare. This
further highlights the extremely low noise of our HSPS.\\
The main performance limitation, i.e. the lower bound for $\alpha$ that we achieved,
is due to the slow rise/fall time of our pulse generator
($\approx2.5$ ns, as seen in the inset plot of Fig.
\ref{peak-out}) and the jitter of the Si-SPAD in
the heralding arm ($\approx 500$ ps). Each of these limits how many non heralded photons can be rejected. The first by providing a minimum width of the switch open time and the second by adding uncertainty in the time between the opening of the switch (driven by the
heralding events) and the presence of the heralded photon. This jitter is
directly related to the spreading of the true coincidences peak
(True) and clearly, $\Delta t_{\textrm{switch}}$ must be kept larger than the
full width of the peak itself (currently $\approx$ 3 ns). \\
These are technical rather than fundamental issues that can be overcome by using lower jitter
commercially available Si-SPADs along with faster pulse generators leading to a possible $\Delta
t_{\textrm{switch}}\lesssim 1$ ns. From the linear trend in Figs. \ref{ONF-plot} and
\ref{alfa-plot} we would expect, at that peak width, the values of $ONF$ to be within 1 \% of zero and $\alpha$ to be within 0.02 of zero at the 1~ns widths that we believe are achievable.

To investigate the extinction performance of our OS, we compare the heralded photon peak within the
switch on-time as in Fig. \ref{peak-in} with the corresponding residual peak of the heralded photon when it is
out of the switch pulse duration as in Fig. \ref{peak-out}, defined as the ratio between
$P^\textrm{(True)}$ in peak-out and peak-in configurations:
\begin{equation}\label{r}
    r=\frac{P^\mathrm{(True)}_{\textrm{peak-out}}}{P^\mathrm{(True)}_{\textrm{peak-in}}} .
\end{equation}
Our measured $r$ value of $3.5\times10^{-3}$ means that our source produces background counts at a
rate of just 37 Hz, this value is comparable to the lowest
dark count rate of the best Si-SPADs currently available. \\
The calibration of our detection apparatus, composed of the fiber beam splitter and the two InGaAs
SPADs, is made using a power-stabilized 1550 nm laser beam attenuated to the photon counting
regime, giving an overall detection efficiency $\eta=(8.1\pm0.2)\%$ \cite{nota2}. This calibration
allows us to evaluate the coupling efficiency $\gamma$ of our single-photon source, defined as:
    \begin{equation}\label{gamma}
    \gamma=\frac{P_{1}^{\mathrm{(True)}}+P_{2}^{\mathrm{(True)}}}{\eta} .
    \end{equation}
The average of the coupling efficiencies obtained for each OS
configuration is $\gamma=(14\pm1)\%$, and the singles measurements
are independent from $\Delta t_{\textrm{switch}}$. We emphasize that better engineering could increase $\gamma$ significantly \cite{coupling2}.\\

In conclusion, we have presented an experimental implementation of a
low-noise heralded single-photon source. The results obtained in
terms of the single-photon parameters $\alpha$ and $ONF$ are
already comparable with the best solid-state based
single-photon sources \cite{shell}. As implemented, $\alpha$ and $ONF$ are limited by the rise/fall
time of the pulse generator controlling the optical
switch and the jitter of the
heralding detector, resulting in a minimum
switch window of a few nanoseconds.

In addition, further improvements in $\alpha$
and $ONF$ are expected with readily available components such as a detector with less than 100 ps
jitter and an optical switch with sub-ns switching times. We note that the inherent switching time
of the optical switch used was 18 GHz.

We also note that, with respect to the other single-photon sources such as for example quantum
dots, color centers in nanodiamond, etc. \cite{shell}, the low-noise heralded-single-photon source
has the advantage of wide wavelength tunability typical of PDC-based sources. Furthermore, because
this source operates at telecom wavelengths, it can exploit commercially available telecom
components, e.g., wavelength division multiplexing and/or narrow spectral selection by means of
Bragg fiber filters.

The background photon rejection and the possibility of controlling and tuning the value of
$t_{\textrm{dead}}$  is particularly advantageous when dealing with slow response systems or slow
detectors, such as for example, detectors with high temporal jitter or slow temporal response
like transition edge superconducting microbolometers \cite{TES}, where the slow response does not allow for temporal discrimination of unwanted events and thus the system is forced to operate at impractically low data rates.

Finally, the whole system can be integrated, or at least
pigtailed, as the source can be realized with a PPLN waveguide, the same technology used for
the fast OS.

\emph{Acknowledgements.} This work has been partially supported by PRIN 2007FYETBY (CCQOTS).



\end{document}